\documentclass[12pt,twoside]{article}
\usepackage{fleqn,espcrc1}

\usepackage{psfig}
\usepackage[figuresright]{rotating}

\newcommand{\AmS}{{\protect\the\textfont2
  A\kern-.1667em\lower.5ex\hbox{M}\kern-.125emS}}

\title{SU(3) Chiral approach to meson and baryon dynamics}

\author{E. Oset\address{Departamento de F\'{\i}sica Te\'orica and IFIC,
Valencia,
	Spain}$^\mathrm{,g}$,
	A. Hosaka\address{Numazu College of Technology, Numazu, Japan},
	J.C. Nacher$^\mathrm{a,g}$,
	M. Oka\address{Tokyo Institute of Technology, Tokyo, Japan},
        J.A. Oller$^a$,
	A. Parre\~no\address{Institute for Nuclear Theory, Washington, Seattle,
	USA},
	J.R. Pel\'aez\address{Departamento de F\'{\i}sica Te\'orica, Madrid,
	Spain},
	A. Ramos\address{Departament d'Estructura i Constituents de la Mat\`eria,
	Barcelona, Spain},
	H. Toki\address{Research Center for Nuclear Physics (RCNP), Osaka,
	Japan}}
       
\begin{document}

\maketitle

\begin{abstract}
We report on recent progress on the 
chiral unitary approach, which is shown to have a much larger convergence radius
than ordinary chiral perturbation theory, allowing one to reproduce data for
meson meson interaction up to 1.2 GeV and meson baryon interaction up to the
first baryonic resonances. Applications to physical processes so
far unsuited for a standard chiral perturbative approach are presented, 
concretely the $K^- p\rightarrow\Lambda(1405)\gamma$ reaction and the $N^\ast
(1535)N^\ast(1535)\pi$ and $\eta$ couplings.
\end{abstract}
\vspace{-0.2cm}
\section{CHIRAL UNITARY APPROACH}

Chiral perturbation theory ($\chi PT$) has proved to be a  very suitable
instrument to implement the basic dynamics and symmetries of the meson meson
and meson baryon interaction \cite{gasser} at low energies. The essence of the
perturbative technique, however, precludes the possibility of tackling problems
where resonances appear, hence limiting tremendously the realm of
applicability. The method that we expose leads naturally to low lying
resonances and allows one to face many problems so far intractable within
$\chi PT$.

The method incorporates new elements: 1) Unitarity is implemented exactly;
2) It can deal with the allowed coupled channels formed by 
pairs of particles from the 
octets of stable
pseudoscalar mesons and ($\frac{1}{2}^+$) baryons; 3) A chiral
expansion in powers of the external four-momentum of the lightest 
pseudoscalars
is done for
Re $T^{-1}$, instead of the $T$ matrix itself which is the case in standard $\chi
PT$.

Within this scheme, and expanding $T_2 ReT^{-1} T_2$ up order $O(p^4)$, 
where $T_2$ is the $O(p^2)$ amplitude, one obtains the matrix
relation in coupled channels 
\begin{equation}
\label{iam}
T = T_2[T_2 - T_4]^{-1} T_2 , 
\end{equation}
where 
 $T_4$ is the usual $O(p^4)$ amplitude in $\chi PT$.

        Once this point is reached one has several options to proceed in 
	decreasing order of complexity:

a) A full calculation of $T_4$ within the same renormalization scheme as in
$\chi PT$ can be done. The eight $L_i$ coefficients from $L^{(4)}$ are then fitted
to the existing meson meson data on phase shifts and inelasticities up to 1.2 GeV, where
4 meson states are still unimportant. This procedure has been carried out in
\cite{dob,gue} in the cases where the complete $O(p^4)$ amplitude has been
calculated. The resulting $L_i$ parameters are compatible with those used in $\chi PT$. 
At low energies the $O(p^4)$ expansion for $T$ of eq. (1) is identical to that
in $\chi PT$. However, at higher energies the nonperturbative structure of eq.
(1),
which implements unitarity exactly, allows one to extend the information
 contained in the chiral Lagrangians to much higher energy than in ordinary 
$\chi$ PT, which is up to about $\sqrt{s}\simeq 400 $ MeV. Indeed it
reproduces the resonances present in the L = 0, 1 partial waves.

b) A technically simpler and equally successful additional approximation
 is generated by ignoring the crossed channel loops and
tadpoles and reabsorbing them in the $L_i$
coefficients given the weak structure of these terms in the physical region.
The fit to the data with the new $\hat{L}_i$ coefficients reproduces the whole meson
meson sector, with the position, widths and partial decay widths of the
$f_0(980)$, $a_0(980)$, $\kappa(900)$, $\rho(770)$, $K^\ast(900)$ resonances in good
agreement with experiment \cite{oller1}. A cut off regularization is used in \cite{oller1} for the
loops in the s-channel. 

c) For the L = 0 sector (also in L = 0, S = $-1$ in the meson baryon interaction)
a further technical simplification is possible. In these cases it is possible
to choose  the cut off such that,  
$\hbox{Re} T_4= T_2 \,\hbox{Re}\, G\, T_2$ where G is the loop function of two
meson propagators. This is possible in those cases
because of the predominant role played by the unitarization of the lowest
order $\chi PT$ amplitude, which by itself leads to the low lying resonances,
 and because other genuine QCD resonances appear at higher energies.

 In such a case and given the fact that $Im T_4 = T_2 Im G T_2$, eq. (1) becomes 
$T = T_2 \,[T_2 - T_2\, G \,T_2]^{-1} \,T_2 = [1 - T_2 \,G]^{-1}\, T_2
\rightarrow T = T_2   + T_2 \,G \,T,$
which is a Bethe-Salpeter equation with $T_2$ and $T$ factorized on shell outside
the loop integral, with $T_2$ playing the role of the potential. This option has
proved to be successful in the L = 0 meson meson sector in \cite{oller2} and in the
L = 0, S = $-1$ meson baryon sector in \cite{osetra}.

       In the meson baryon sector with S = 0, given the disparity of the
        masses in the coupled channels $\pi N$, $\eta N$, $K\Sigma$,
        $K\Lambda$,
        the simple ``one cut off approach'' is not possible. In \cite{kaiser} higher
        order Lagrangians are introduced while in \cite{par} different subtraction
        constants in G are incorporated
        in each of the former channels leading in both cases to acceptable
        solutions when compared with the data.

 An alternative, related procedure to eq. (\ref{iam}), is developed in 
 \cite{olleroset} using the $N/D$
method and allowing the contribution of preexisting mesons which remain in the
limit of large $N_c$. This procedure allows one to separate the physical mesons
into preexisting ones, mostly $q\bar{q}$ pairs and the others which come as
resonances of the meson meson scattering due to unitarization.
\vspace{-0.6cm}
\section{APPLICATION TO THE $K^- p\rightarrow\Lambda(1405)\gamma$ REACTION}

Using the option c), a good description of interaction of the $K^-p$ is coupled
channels interaction 
is obtained in terms of the lowest order Lagrangian and the Bethe Salpeter
equation with a single cut off. One of the interesting features of the approach
is the dynamical generation of the $\Lambda(1405)$ resonance just below the
$K^-p$ threshold. The threshold behavior of the $K^-p$ amplitude is thus
very much tied to the properties of this resonance. Modifications of these
properties in a nuclear medium can substantially alter the $K^-p$ and $K^-$
nucleus interaction and experiments looking for these properties are most welcome. 

        In a recent paper \cite{nac} we propose the $K^- p\rightarrow
        \Lambda(1405)\gamma$ reaction as a means to study the
        resonance, together with the $K^- A\rightarrow
        \Lambda(1405)\gamma A'$ reaction to see the modification of its properties
        in nuclei. The $\Lambda(1405)$ is seen in its decay products
        in the $\pi\Sigma$ channel, but as shown in \cite{nac} the sum of the cross
        sections for $\pi^0\Sigma^0$, $\pi^+\Sigma^-$, $\pi^-\Sigma^+$
        production has the shape of the resonance $\Lambda(1405)$ in the I = 0
        channel. Hence, the detection of the $\gamma$ in the elementary reaction,
         looking at $d\sigma/dM_I$ ($M_I$ being the invariant mass of the meson
         baryon system which can be obtained from the $\gamma$ momentum), is
         sufficient to get a clear $\Lambda(1045)$ signal. In fig. 1 we show the cross
         sections predicted for the $K^- p\rightarrow\Lambda(1405)\gamma$
         reaction looking at $\gamma\pi^0\Sigma^0$, $\gamma$ $all$ and $\gamma \Lambda(1405)$
         (alone). All of them have approximately the same shape and strength
         given
         the fact that the I = 1 contribution is rather small. The momentum chosen for the $K^-$ is 500 MeV/c which makes it
  suitable of experimentation at KEK and others facilities.
\begin{figure}[h]
\centerline{\protect
\hbox{
\psfig{file=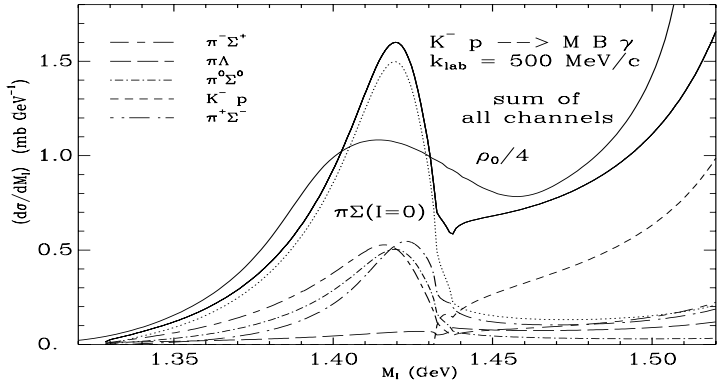,height=5cm,width=10cm,angle=-90}}}
\caption{Mass distribution for the different channels, as a
function of the invariant mass $M_I$ of the final meson baryon system.}
\end{figure}
\vspace{-1.1cm}
\section{$N^\ast(1535)N^\ast(1535)\pi,\eta$ COUPLINGS}

Since the $N^\ast$(1535) resonance is also obtained via the Bethe-Salpeter
equation (see fig. 2) in the meson baryon $S=0$ sector with the channels
$\pi N$, $\eta N$, $K\Sigma$, $K\Lambda$, one can automatically generate
the series implicit in fig. 3 which provides the coupling of a $\pi$ or an
$\eta$ to the resonance, the latter being generated at both sides of the
external mesonic vertex. All vertices needed for the calculation can be obtained
from standard chiral Lagrangians. The results which we obtain are \cite{hosaka1}
$\frac{g_{\pi^0 N^\ast N^\ast}}{g_{\pi^0 N N}} = 1.3  ; 
\frac{g_{\eta N^\ast N^\ast}}{g_{\eta N N}} = 2.2$
\begin{figure}[h]
\centerline{\protect
\hbox{
\psfig{file=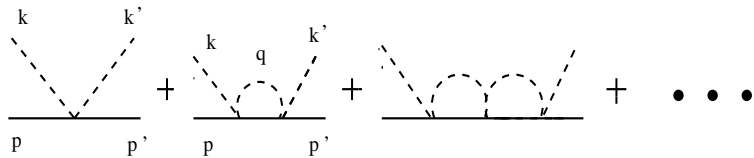,height=2cm,width=10cm,angle=-90}}}
\vspace{-0.9cm}
\caption{Diagrammatic representation of the Bethe Salpeter equation}
\end{figure}
The result for the $\pi$ coupling rule out the mirror assignment in chiral
models where the nucleon and the $N^\ast(1535)$ form a parity doublet
\cite{hosaka1,hosaka2}
in analogy with the linear $\sigma$ model.
\begin{figure}[h]
\centerline{\protect
\hbox{
\psfig{file=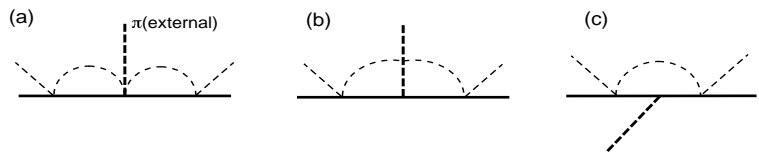,height=2cm,width=10cm,angle=0}}}
\vspace{-0.9cm}
\caption{Diagrammatic representation of the $\pi N^\ast N^\ast$ coupling}
\end{figure}
\vspace{-0.3cm}
\section{ SUMMARY}
  We have reported on the unitary approach to meson meson and meson baryon
  interactions using chiral Lagrangians, which has proved to be an efficient
  method to extend the information on chiral symmetry breaking to higher
  energies where $\chi PT$ cannot be used. This new approach has opened the
  doors to the investigation of many new problems so far intractable with $\chi PT$
  and a few examples have been reported here. We have applied these techniques
  to the $K^- p\rightarrow\Lambda(1405)\gamma$ reaction and the evaluation
  of the $N^\ast N^\ast\pi,\eta$ couplings. The experimental implementation of the
  former reaction and others on photoproduction of scalar mesons and of the 
  $\Lambda(1405)$, reported elsewhere \cite{grana}, will provide new
  tests of these emerging pictures implementing chiral symmetry and unitarity.
  Similarly, the techniques used to evaluate the $N^\ast N^\ast\pi,
  \eta$ couplings can be easily extended to evaluate electromagnetic properties
  of low lying resonances which are generated within the unitary scheme.

\end{document}